\documentclass[5p, times, authoryear]{elsarticle}

\usepackage{lineno}
\modulolinenumbers[5]
\usepackage{graphicx}
\usepackage{caption}
\usepackage{subcaption}
\usepackage{float}
\usepackage{amsmath}
\usepackage{hyperref}
\usepackage{cleveref}
\usepackage[authoryear]{natbib}

\journal{Education for Chemical Engineers}

\biboptions{authoryear}

\begin{document}

\begin{frontmatter}

\title{Introducing students to research codes: A short course on solving partial differential equations in Python }
% \tnotetext[mytitlenote]{Fully documented templates are available in the elsarticle package on \href{http://www.ctan.org/tex-archive/macros/latex/contrib/elsarticle}{CTAN}.}

\author[First,Second]{Pavan Inguva\corref{mycorrespondingauthor}}
\cortext[mycorrespondingauthor]{Corresponding author}
\ead{inguvakp@mit.edu}
\author[Second]{Vijesh J. Bhute}
\author[Second]{Thomas N.H. Cheng}
\author[Second]{Pierre J. Walker}
\address[First]{Department of Chemical Engineering, Massachusetts Institute of Technology, 25 Ames Street, Cambridge, MA 02142, United States}
\address[Second]{Department of Chemical Engineering, Imperial College London, London SW7 2AZ, United Kingdom}

%\fnref{myfootnote}}

% \fntext[myfootnote]{Since 1880.}

% \author[mymainaddress,mysecondaryaddress]{Elsevier Inc}

% \author[mysecondaryaddress]{Global Customer Service\corref{mycorrespondingauthor}}
% \cortext[mycorrespondingauthor]{Corresponding author}

% \address[mymainaddress]{1600 John F Kennedy Boulevard, Philadelphia}
% \address[mysecondaryaddress]{360 Park Avenue South, New York}

\begin{abstract}
Recent releases of open-source research codes and solvers for numerically solving partial differential equations in Python present a great opportunity for educators to integrate these codes into the classroom in a variety of ways. The ease with which a problem can be implemented and solved using these codes reduce the barrier to entry for users. We demonstrate how one of these codes, FiPy, can be introduced to students through a short course using progression as the guiding philosophy. Four exercises of increasing complexity were developed. Basic concepts from more advanced numerical methods courses are also introduced at appropriate points. To further engage students, we demonstrate how an open research problem can be readily implemented and also incorporate the use of ParaView to post-process their results. Student engagement and learning outcomes were evaluated through a pre and post-course survey and a focus group discussion. Students broadly found the course to be engaging and useful with the ability to easily visualise the solution to PDEs being greatly valued. Due to the introductory nature of the course, due care in terms of set-up and the design of learning activities during the course is essential. This course, if integrated with appropriate level of support, can encourage students to use the provided codes and improve their understanding of concepts used in numerical analysis and PDEs. 
\end{abstract}

\begin{keyword}
Partial differential equation\sep Mathematics \sep Python \sep Introductory \sep Experiential learning \sep FiPy
% \MSC[2010] 00-01\sep  99-00
\end{keyword}

\end{frontmatter}

% \linenumbers

\section{Introduction}

Partial differential equations (PDEs) are of significant interest to scientists and engineers as they can be used to study a variety of physical and engineering problems such as transport processes \citep{Bird2002}, structural mechanics \citep{Zien2005} and electromagnetism, quantum physics \citep{Gowers2008} to list a few. For engineers, PDEs are often one of the most complex mathematical topics that would be covered in the undergraduate course. However, solving PDEs for systems of academic and industrial interest often require numerical methods as analytical solutions may not be available. Therefore, subsequent courses on PDEs for engineers often focus on applications in different settings and employ numerical methods and/or empirical correlations to obtain usable results. 

The importance of numerical methods and the development of solver codes is evident in the sheer diversity and sophistication of PDE solver packages and codes currently available. In addition, there is a large variety of ancilliary elements and tools such as pre and postprocessing softwares, wrappers and webapps with varying levels of functionalities. These codes and software are either available commercially or open-source or they can be developed in-house to meet specific needs. Educators face a complex challenge in deciding which set of tools should be used in their classroom. Multiple factors such as scope and objectives of the course, students' background and availability of resources need to be considered and balanced. We propose the following factors to consider when evaluating a tool: 
\begin{enumerate}
    \item Availability: Open-source software is much more available to users compared to commercial software which can be quite costly. 
    \item Ease of use: Factors such as the learning curve associated with using the software and/or the availability of reference material and support should be considered. 
    \item Scalability: Can the software be used for both smaller tasks such as testing and teaching and also be used for large problems in industrial and research settings. 
    \item Learning objectives: Is the class more theoretical and focused on the mathematics and physics of the problem and/or numerical method or is the focus on the application of a tool to solve an engineering problem. Another factor to consider is the academic background of the students e.g. is the course aimed at an undergraduate or graduate level. 
\end{enumerate}

To date, no code or software has been demonstrated to effectively address all four factors listed above in a teaching setting. Commercial software such as ANSYS and COMSOL are commonly used by industry and academics and scale very well. Furthermore, there is comprehensive reference material available and a variety of courses at the undergraduate and graduate level utilize these software, often in an application setting \citep{Mors2013,Vicens2014,Adair2014,Huang2019, Wica2020}. However, these commercial packages fall short in terms of availability due to their high costs \citep{Chen2014} and as many of the numerical complexities are abstracted away, they are not readily usable for teaching the fundamentals aspects of solving PDEs numerically.

MATLAB, whilst commercial in nature and therefore restricted in access, is often used by educators in numerical methods courses \citep{Kiu2015}. However, such student generated codes scale poorly as they are meant to be instructive in nature. These courses are also often at a senior undergraduate or graduate level which restricts the target audience. Educators also commonly use MATLAB and other similar commercially available software in application settings for problems involving systems of ordinary differential equations (ODEs) \citep{Ibrahim2010,Li2016,Molina2019} or PDEs \citep{Stephens2019}. 

Fortunately, for many systems involving ODEs, including chemical kinetics, process modelling or complex systems, there are a wide variety of open-source libraries in Python such as SciPy \citep{Vir2020} or other languages such as Julia \citep{Rack2017} which can be used effectively. These libraries and Python in general have found significant use in both research and teaching settings \citep{Golman2016,Wang2017,Sawaki2020}. 

However, the open-source landscape for solving PDEs is a bit more nuanced. Many open-source codes which are incredibly powerful and have been gaining traction in academia and industry such as OpenFOAM \citep{Jasak2007} or FEniCS \citep{Fenics2011} have effectively addressed issues such as availability and scalability. However, care must be taken with regards to using these codes in the classroom as factors such as the ease of use may adversely impact meeting module learning objectives. For example, OpenFOAM is well known for having a steep learning curve \citep{Chen2014} and is therefore ill-suited for introductory courses on numerical methods or in an application setting. 

We believe that the recent development of a variety of open-source PDE solver codes in Python presents an incredible opportunity for educators. The following list of solver codes is non-exhaustive and is categorised by discretisation method:
\begin{description}
\item[Finite volume method] FiPy \citep{Guyer2009}
\item[Finite element method] FEniCS \citep{Fenics2011}, SfePy \citep{Sfepy2019}
\item[Finite difference method] Scikit FiniteDiff \citep{Cellier2019}, Pystencils \citep{Bauer2016}
\item[Spectral methods] Dedalus \citep{Burns2019}

\end{description}

These solver codes can effectively address all four considerations. Being open-source means that these solvers are easily accessible to users. With the front-end written in Python, users are able to quickly understand and write code to solve problems of interest \citep{Guyer2009}. In addition, most of these solvers have active user discussion forums or means to get support from the developers and have comprehensive documentation available. Many of these codes are also actively being used in research and have been demonstrated to effectively scale \citep{Fenics2011,Burns2019,Sfepy2019}. Lastly, when collectively considered as a set of tools, these solver codes contain a wide variety of features and capabilities that can be used by users of varying skill level, giving educators a remarkable design space when considering integrating these codes into their course. 

As a first iteration, we demonstrate the use of one of these codes, FiPy, as part of an introductory course to numerically solving PDEs with Python. A short course which ran over two days was developed and rolled out over the summer of 2020. We hope to integrate parts of this course into an advanced engineering mathematics module to complement the teaching of the PDE section of the course. The extensive course notes and all code used in the course will be made publicly available. We intend to supplement the code repository in due course with additional contributions to expand the material offered.

\section{Course context and structure}

The course outlined in the present work is rolled out as a course run over two three-hour sessions held on two days in the summer of 2020. We intend to integrate components of the course into the PDE section of the second year Math II (Engineering Mathematics) module that students are required to take. The Math II module is quite comprehensive, covering topics such as vector calculus, Fourier series, PDEs and statistics. Prior to this module, students would have taken a first year mathematics module covering univariable calculus, linear algebra and ordinary differential equations along with an introductory coding course in MATLAB. Students would have also completed an introductory class on transport processes. Transport phenomena is chosen as the context for most of the exercises as students would already have a conceptual understanding of the problems. Educators in different disciplines can readily implement problems of interest to them with comparative ease. 

To help students appreciate the whole pipeline of setting up a simulation and analysing the results, the course starts with a quick guide on how to set up Python and install FiPy. Four exercises of increasing complexity are developed for students to explore. Each exercise may be further broken down into parts to emphasise specific points of interest. Each exercise was designed to simultaneously introduce the necessary content at that stage and also lay the groundwork for students to progress to the next exercise to enable students to progress to more complex problems \citep{Xie2020}. In exercise 1, students explore the Laplace equation in a problem from their module while in exercise 4, students consider the Cahn--Hilliard equation to model polymer blend demixing. Exercise 4 also introduces students to using ParaView \citep{Paraview2020}, a research-grade software often used by researchers to analyse and post-process simulation data.

As students may not have experience with Python and may be intimidated by the idea of numerically solving PDEs, much care was taken to ensure that the course is as accessible as possible. A detailed step-by-step guide is provided to set up the environment and install relevant software before the beginning of the course in a pre-course document. All code is written with extensive comments explaining what each line of code means. It is desired that students would see the code provided to them as a reference for them to use in the future. Furthermore, comprehensive notes with annotated screenshots were made for all parts of the course. The notes and other course material were made in collaboration with students to improve accessibility \citep{Inguva2018,Xie2020}. 

A pre and post course survey was designed and given to participants in the course to fill out anonymously. The survey was designed to evaluate students' perception of the course structure and approach and their self-efficacy and attitudes before and after the course. A series of statements were provided and students were asked to rate how strongly they agreed with the statements. A focus group discussion was also conducted to get students' feedback on how they perceived the course and also to garner feedback and suggestions on future strategies for course improvement and integration into the classroom. 

\section{Set-up, exercises and post-processing}

\subsection{Python environment}

We used the Spyder Integrated Developer Environment (IDE) which is distributed with the Windows Installation of Anaconda \citep{Anaconda2016}. Students are also given instructions on how to create an Anaconda environment as a tool for creating reproducible environments for simulations. The associated environment files for Windows and Linux installations are available in the code repository. 

\subsection{Exercises}

\subsubsection{Exercise 1: Laplace equation}

The first exercise serves as an introduction to working with FiPy and basic concepts in solving PDEs. The exercise aims to achieve the following:
\begin{enumerate}
    \item Set up a simple case 
    \item Run the simulation
    \item Visualise the results
    \item Benchmark the results of the simulation to the analytical solution
\end{enumerate}

The classic 2D Laplace equation on a unit square domain is used as the equation to be solved with the following boundary conditions: 
\begin{equation}
    \nabla ^{2} \phi =0, 
\end{equation}
\begin{equation}
    \phi (0,y) = \phi(1,y) = \phi(x,0) = 0,\phi (x,1) = 1,
\end{equation}

where $\phi$ is the solution variable and $x,y$ are the $x$ and $y$ coordinates, respectively. The analytical solution of the problem is written as follows: 
\begin{equation}
    \phi(x,y) = \sum_{n=0}^{\infty} \frac{4 \sin ( (2n+1) \pi x )\sinh( (2n+1) \pi y )}{(2n+1) \pi  \sinh ({(2n+1) \pi})}. 
\end{equation}

\begin{figure*}[htbp]
    \centering
    \begin{subfigure}[t]{0.45\linewidth}
		\centering
		\includegraphics[height=9cm, width=9cm]{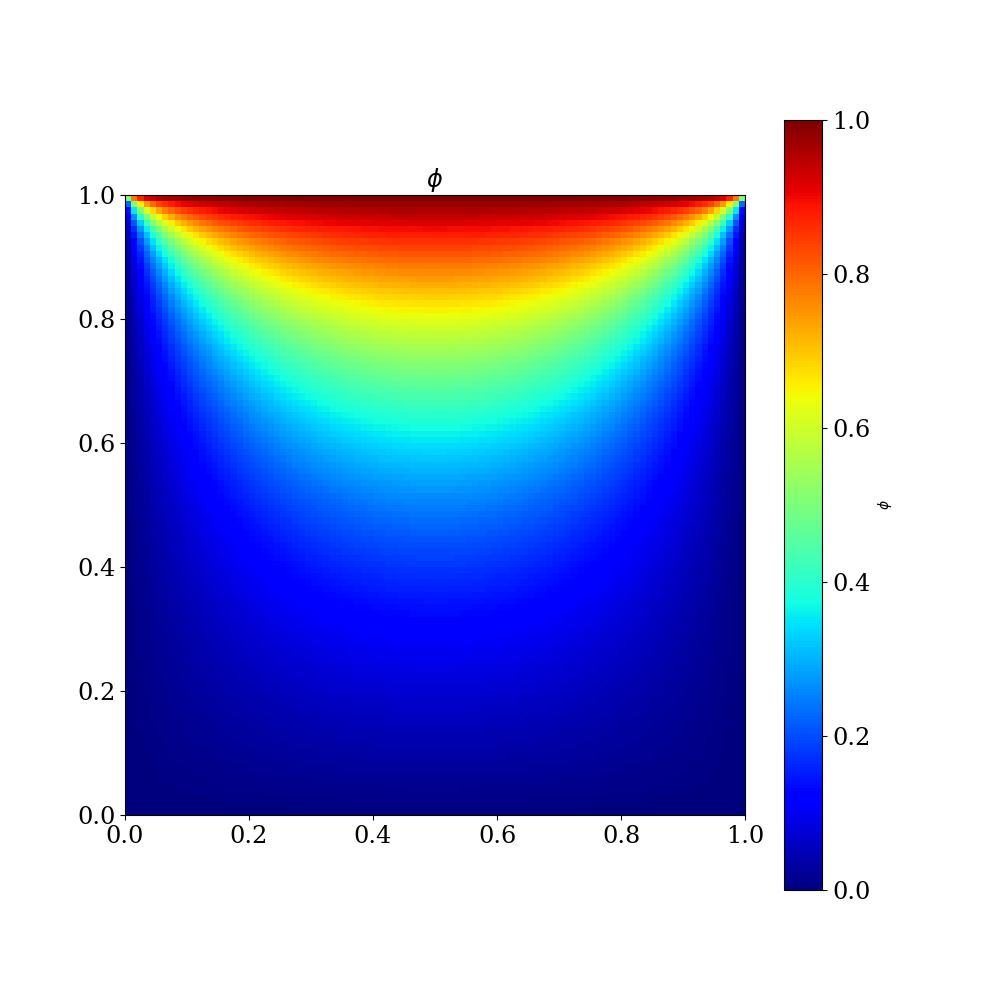}
		\caption{Numerical solution}
	\end{subfigure}
	\quad
	\begin{subfigure}[t]{0.45\linewidth}
		\centering
		\includegraphics[height=9cm, width=9cm]{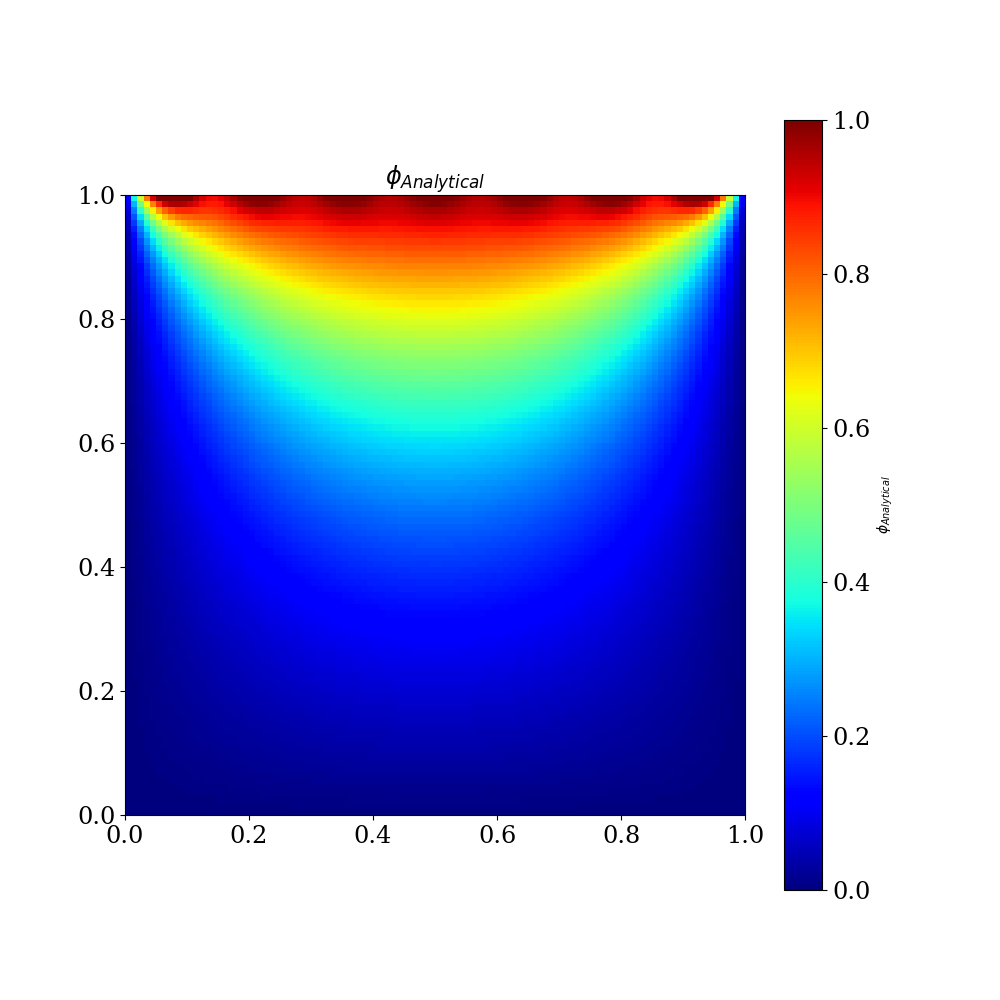}
		\caption{Analytical solution truncated to the 7th term}
	\end{subfigure}
	\quad
	\caption{Numerical (a) vs analytical solution (b) to exercise 1 generated using the default FiPy viewer.}
	\label{fig:ex1}
\end{figure*}

\subsubsection{Exercise 1: Results}
As shown in \cref{fig:ex1}, the numerical and analytical solutions to the problem posed in exercise 1 are highly comparable. However, the impact of the truncation of the analytical solution can be seen whereby the solution at the top vertex is less smooth compared to the numerical solution. These figures are generated using the default FiPy viewer and students would be able to see these results as they execute the scripts in the IDE. 

To help explain the results observed, the idea of steady-state conduction/diffusion on a plate was used to provide a physical intuition of the results. The vibrant colours of the result also helped facilitate introducing the idea of contour / isolines as lines of constant $\phi$ can be visually identified quite easily.  

\subsubsection{Exercise 2: Diffusion equation}

The second exercise introduces transient problems in the form of the familiar diffusion equation without or with a reaction. The exercise is broken down into three parts: In part 2A, the diffusion equation is solved with Dirichlet boundary conditions while in part 2B, Neumann boundary conditions are applied. In the last part 2C, a reaction term is added to the diffusion equation. All the problems are solved in 1D. Exercise 2 aims to achieve the following: 

\begin{enumerate}
    \item Introducing how time stepping is carried out and how different time stepping algorithms (implicit vs explicit) can affect the solution.
    \item Appreciate how the numerical solution compares with transient analytical solution. 
    \item Introduce the idea of a source / sink term and how it can be used to model a reaction. 
\end{enumerate}

The problem statements for each part are as follows: 

\paragraph{Part 2A} 
\begin{equation}
    \label{eq:simplediffusion}
    \frac{\partial c}{\partial t} = D \nabla^{2}c,
\end{equation}
\begin{equation}
    c(x=0, t) = 1,c(x=1, t) = 0,c(x, t=0) = 0,
\end{equation}
where $c$ is the concentration, $t$ is the time and $D$ is the diffusion coefficient which is set to $1.0$. This problem has an analytical solution which can be written as follows:
\begin{equation}
    c(x,t) = 1 - x - \sum_{n=1}^{\infty}\frac{2}{n \pi} \sin{(n\pi x) e^{-Dn^{2}\pi^{2} t}}.
\end{equation}
The mesh Fourier number $\mathrm{Fo}_{\mathrm{Mesh}}$ is introduced as a stability criteria for explicit forward Euler time-stepping. As a starting point, students are provided a value of $\Delta t = 0.00018$ which is stable. 
\begin{equation}
    \Delta t \leq \frac{\Delta x ^{2}}{2D}
\end{equation}

\paragraph{Part 2B}
The same equation \cref{eq:simplediffusion} as part 2A is solved, except the initial and boundary conditions are as follows: 
\begin{align}
    \frac{\partial c}{\partial x}(x=0, t) = 0, \frac{\partial c}{\partial x}(x=1, t) = 0, \nonumber \\ 
    c(x, t=0) = 0.5 + 0.3\sin{(2\pi x)}.  
\end{align}

\paragraph{Part 2C} 
A reaction term is introduced to the diffusion equation as follows: 
\begin{equation}
    \frac{\partial c}{\partial t} = D \nabla^{2}c + \kappa c,
\end{equation}
where $\kappa$ is the rate constant and is set to $-2.0$. The initial and boundary conditions are the same as part 2A. The reaction term is treated as an implicit source term. 

\subsubsection{Exercise 2: Results}
\begin{figure*}[htbp]
    \centering
    \begin{subfigure}[t]{0.45\linewidth}
		\centering
		\includegraphics[height=9cm, width=9cm]{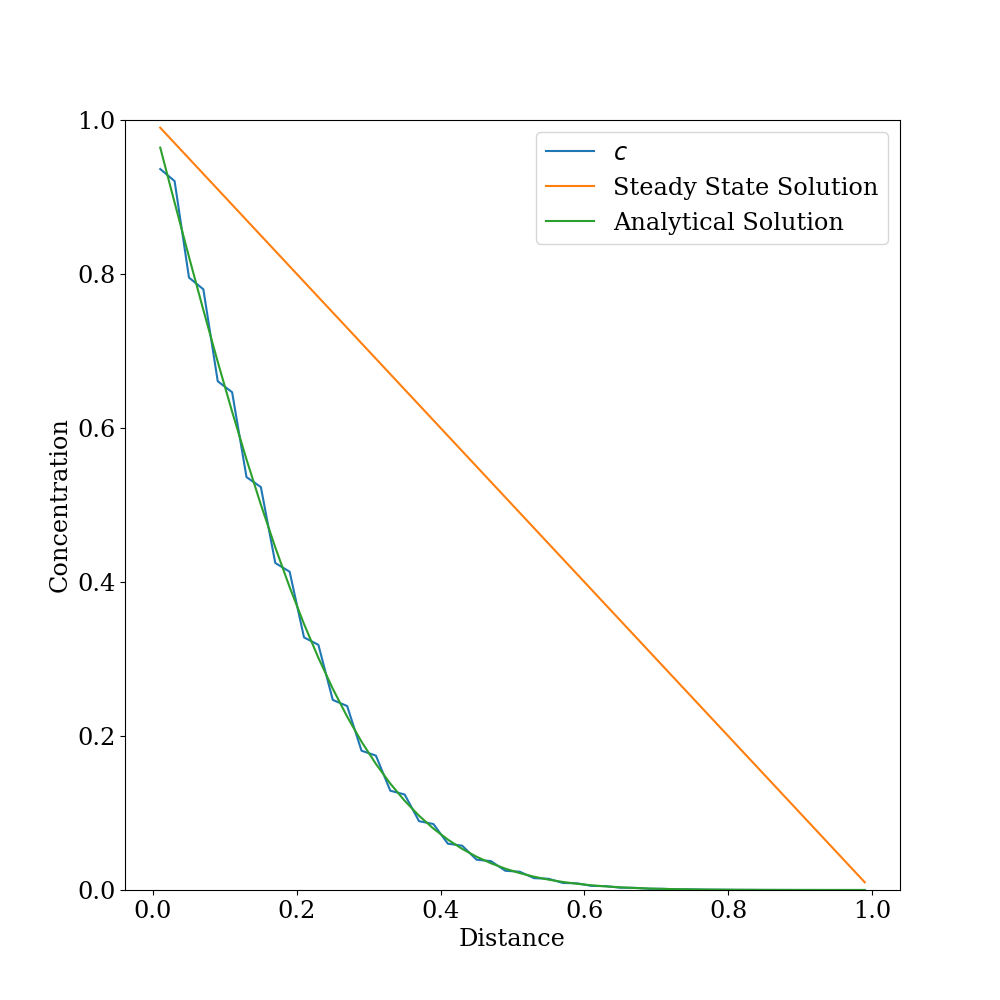}
		\caption{Explicit forward Euler time-stepping with $\Delta t = 0.0001998$}
	\end{subfigure}
	\quad
	\begin{subfigure}[t]{0.45\linewidth}
		\centering
		\includegraphics[height=9cm, width=9cm]{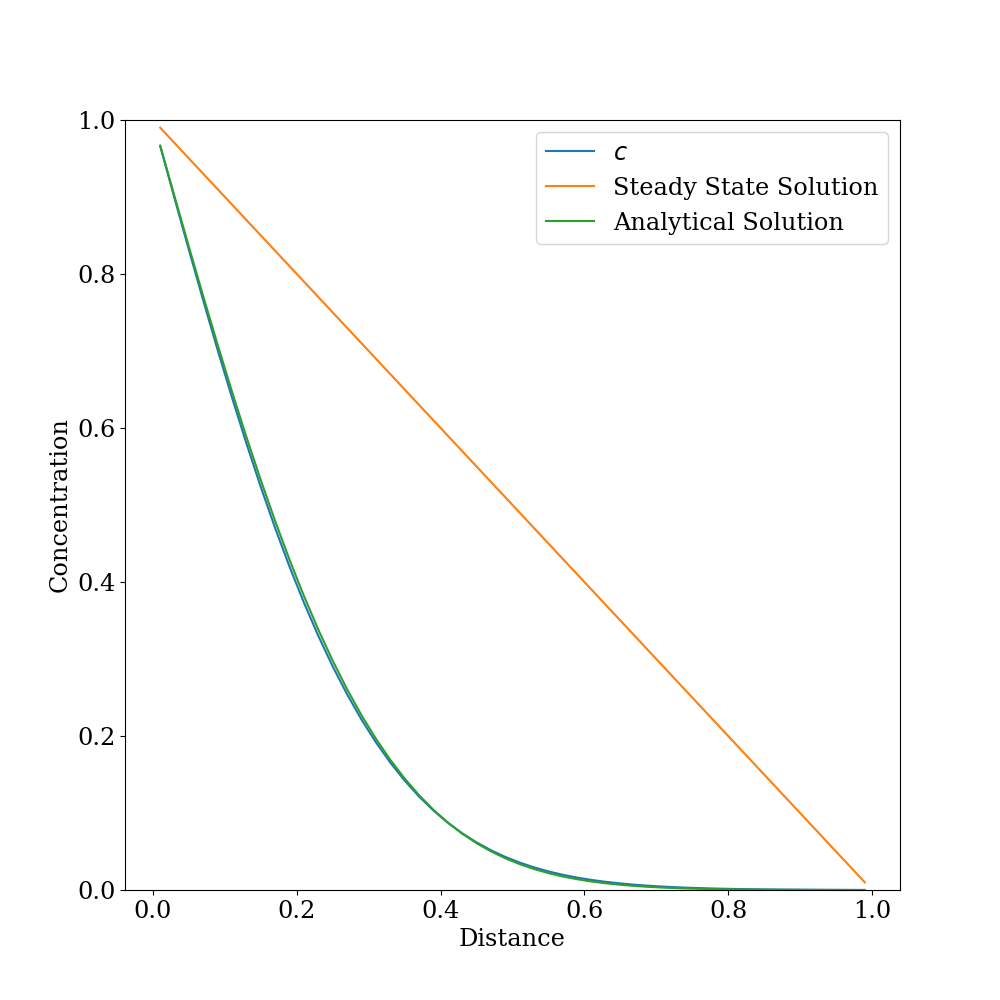}
		\caption{Implicit backward Euler time-stepping with $\Delta t = 0.0018$}
	\end{subfigure}
	\quad
	\caption{Explicit vs implict time-stepping for exercise 2a.Plots are generated using the default FiPy viewer.}
	\label{fig:ex2a}
\end{figure*}

To fully appreciate the results from part 2A, students will need to explore a few settings namely the time-stepping scheme,$\Delta t$ and the time-stride. A time-stride parameter is introduced into the code to allow students to visualise every nth step which is especially important when explicit time-stepping is used due to the small $\Delta t$. When an explicit forward Euler time-stepping scheme is used, the default value of $\Delta t$ provided to students yields a stable solution. However, when $\Delta t = 0.0001998$, numerical instabilities in the form of "wiggles" become noticeable as shown in \cref{fig:ex2a} and further increasing $\Delta t$ causes the unacceptably oscillatory behaviour which is the expected behaviour. However, using an implicit time-stepping scheme enables users to use much higher values of $\Delta t$ and still achieve stable results. 

If students use a small time-step and a small time-stride (such as the given value of the time-step and a time-stride of 1), they would also be able to view the initial profile of the truncated analytical solution which has the effects of the higher frequency Fourier modes visible. As time progresses, the higher frequency modes decay very quickly and the truncated analytical solution corresponds to the numerical solution very well. This can help students to visualise various theoretical concepts pertaining to Fourier series. 

\begin{figure}[htbp]
    \centering
    \includegraphics[width=0.9\linewidth]{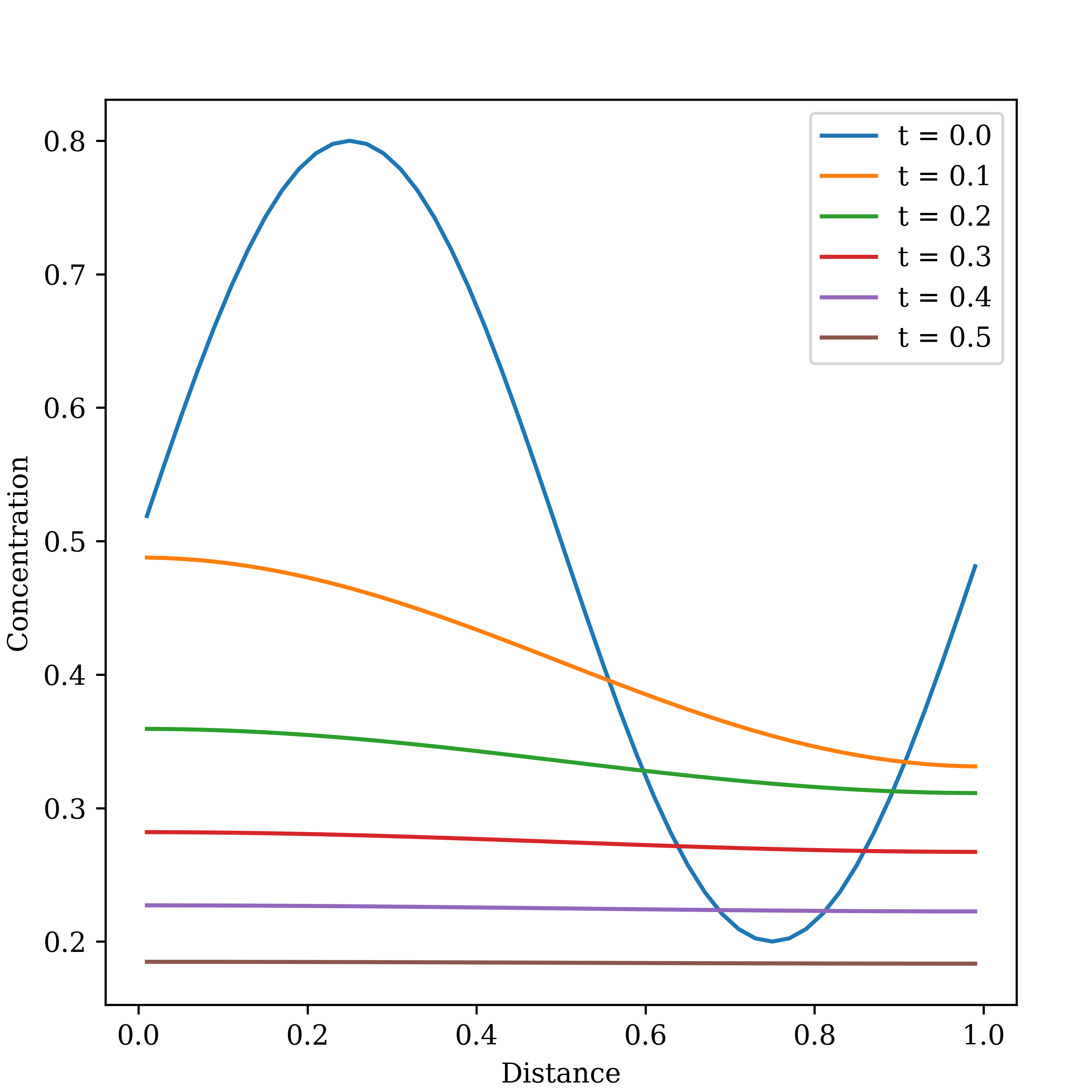}
    \caption{Evolution of concentration for exercise 2C at different times.}
    \label{fig:ex2c}
\end{figure}

Conceptually, parts 2B and 2C are quite similar. We present the expected results from part 2C in \cref{fig:ex2c} to demonstrate a typical plot students can generate if they use the TSVViewer in FiPy to export their data and generate more complicated plots on their own. 

\subsubsection{Exercise 3: Convection-Diffusion-Reaction (CDR) equation}

The CDR equation is of significant use to model the transport of chemical species. Part 3A of the exercise assumes no reaction and serves to visually demonstrate how the convective and diffusive transport function. Concepts such as the discretisation of the convection term and the derivation of the Courant number are covered in the notes. Part 3B considers a coupled system of equations with a non-linear reaction term. The aims of exercise 3 are as follows: 
\begin{enumerate}
    \item Visualise how convective and diffusive transport occur.
    \item Appreciate the significance of Peclet number in convection-diffusion problems
    \item Appreciate the added complexity involved in modelling convection.
    \item Appreciate how scaling a problem facilitates its solution
    \item Implement a coupled equation system with a non-linear source term. 
\end{enumerate}

The problem statement for each part is as follows: 

\paragraph {Part 3A} 
The convection-diffusion equation is first scaled and solved in non-dimensional form as follows: 
\begin{equation}
    \frac{\partial c}{\partial \tilde{t}} + \frac{\partial c }{\partial \tilde{x}} = \frac{1}{\mathrm{Pe}}\frac{\partial^{2} c}{\partial \tilde{x}^{2}},
\end{equation}
where $\tilde{t}$ is the dimensionless time and is scaled by the characteristic length $L$ and velocity $U$ as such $t = \frac{L}{U} \tilde{t}$.$\tilde{x}$ is the dimensionless spatial coordinate and is scaled as such $x = L \tilde{x}$. Lastly, $\mathrm{Pe}$ is the Peclet number and is given by $\mathrm{Pe} = \frac{UL}{D}$. $U$ and $L$ are set to 1.0 and $\mathrm{Pe}$ is set to 100. A 1D mesh of length $4$ was generated with a mesh size $\Delta \tilde{x} = 0.01$. The simulation is run until $t = 2.0$. Periodic boundary conditions are applied:
\begin{equation}
    c(\tilde{x} = 0, \tilde{t}) = c(\tilde{x} = 4 , \tilde{t}).
\end{equation}
The concentration $c$ is initialised with a narrow Gaussian distribution:
\begin{equation}
    c(\tilde{x}, \tilde{t}=0) = \frac{0.05}{0.05 \sqrt{2 \pi}} e^{\frac{-1}{2} \bigg( {\frac{\tilde{x} - 0.5}{0.05}} \bigg)^{2}}.
\end{equation}

\paragraph{Part 3B}
The simple chemical reaction between species A and B to form C which is shown as follows: 
\begin{equation}
    A + B \to C,
\end{equation}
is given as first order with respect to both species $A$ and $B$ with the following simplified expression for the reaction rate: 
\begin{equation}
    r_{A} = r_{B} = -r_{C} = \kappa ab,
\end{equation}
where $r_{i}, i \in {A, B, C}$ is the reaction rate for each of the species A, B and C, $\kappa$ is the rate constant and $a, b, c$ is the concentration of species A, B and C, respectively. The equation system for the three species can be written as follows: 
\begin{align}
    &\frac{\partial a}{\partial t} + U\frac{\partial a}{\partial x} - D\frac{\partial^{2}a}{\partial x^{2}} + r_{A} = 0, \nonumber
\\
    &\frac{\partial b}{\partial t} + U\frac{\partial b}{\partial x} - D\frac{\partial^{2}b}{\partial x^{2}} + r_{B} = 0,
\\
    &\frac{\partial c}{\partial t} + U\frac{\partial c}{\partial x} - D\frac{\partial^{2}c}{\partial x^{2}} + r_{C} = 0, \nonumber
\end{align}
where $U$ and $D$ are the convective velocity and diffusion coefficient respectively and both are assumed to be constant and identical for all species. $D$, $\kappa$ and $U$ are set at $1.0$, $30.0$ and $3.0$, respectively. The simulation domain is set at $L = 4$. Dirichlet boundary conditions are specified at the inlet as follows: 
\begin{align}
    &a(x=0, t) = 1.0, \nonumber \\ 
    &b(x=0, t) = 0.5, \\
    &c(x=0, t) = 0.0. \nonumber
\end{align}
However, specifying the boundary conditions at the outlet is non-trivial. The appropriate boundary condition is termed the "free boundary condition", but is outside the scope of this course. Often, a zero-diffusive flux boundary condition is applied in such cases \citep{Lynch2020} and it suffices for pedagogical purposes. 
\begin{align}
    &\frac{\partial a}{\partial x}(x=L, t) = 0, \nonumber \\
    &\frac{\partial b}{\partial x}(x=L, t) = 0, \\
    &\frac{\partial c}{\partial x}(x=L, t) = 0. \nonumber
\end{align}

This equation system introduces students to two interesting concepts in numerical methods. The first is the idea of strong vs weak coupling. The transport equations for species A and B are strongly coupled and need to be solved at the same time, however the transport equation for species C is weakly coupled and can be lagged in the solution process. The second concept is the idea of "sweeps". As the reaction terms are non-linear, the equation system is non-linear necessitating solving the linear equation system iteratively at that time-step using the result from the previous solution \citep{Guyer2013}. Students may typically solve the steady-state version of this problem as a set of coupled ODEs numerically, but they now have the opportunity to explore the transient dynamics of the solution.

\subsubsection{Exercise 3: Results}
\begin{figure}[htbp]
    \centering
    \includegraphics[width=0.9\linewidth]{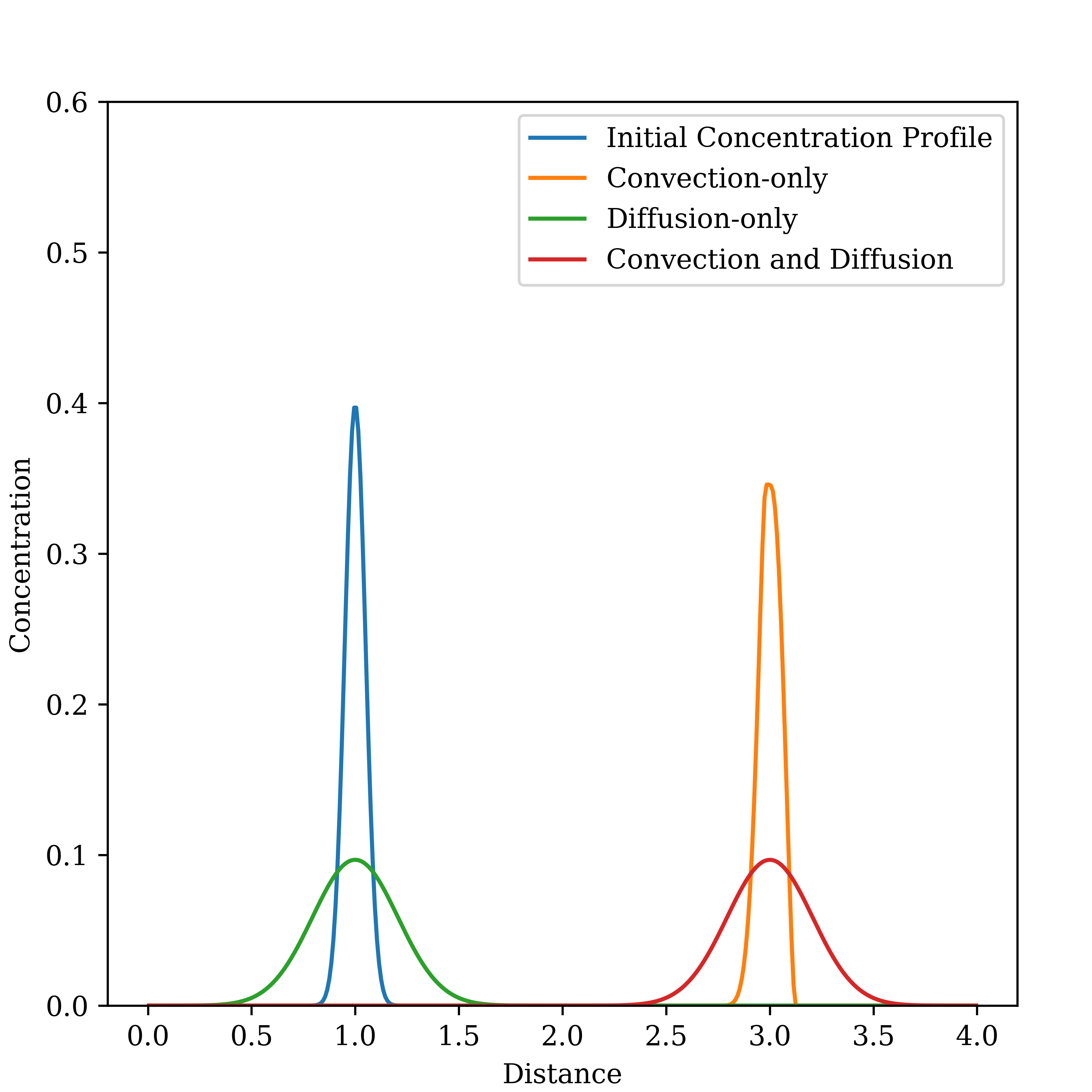}
    \caption{Effect of different transport mechanisms. Concentration profiles @ $t=2.0$ are shown together with the initial concentration profile.}
    \label{fig:ex3a}
\end{figure}

The effects of the different transport mechanisms are shown in \cref{fig:ex3a}. Students are able to solve each equation (convection-only, diffusion-only and convection + diffusion) by commenting out a few lines in the code. The notes introduce the concept of numerical diffusion which is an important challenge in modelling convection and is demonstrated by the false diffusion present in the convection-only system as shown in \cref{fig:ex3a}. FiPy has multiple convection schemes present, and the VanLeer convection scheme was found to have the least amount of numerical diffusion and is therefore used. Students can explore how different parameters such as the mesh resolution and convection scheme can impact the results with minimal modification to the code. 

\begin{figure}[htbp]
    \centering
    \includegraphics[width=0.9\linewidth]{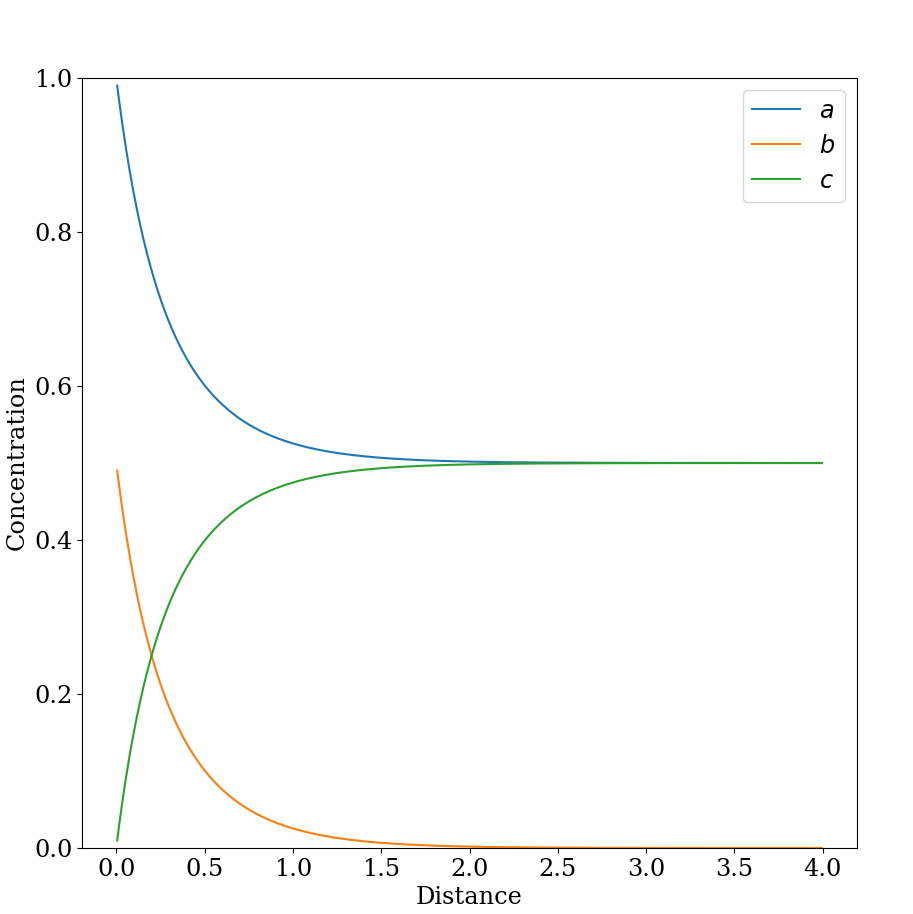}
    \caption{Concentration profiles of species $a,b$ and $c$ @ $t=5.0$. The system appears to have reached a steady-state.}
    \label{fig:ex3b}
\end{figure}

\subsubsection{Exercise 4}

To wrap up the course, we wanted to expose students to a problem that is currently of significant academic interest and is comparatively simple to implement in FiPy. A modified Cahn--Hilliard equation system based on the work of \citet{Petr2012} is selected as the model problem. This equation system can be used to study polymer blend demixing \citep{Inguva2020, Naum1994} and can be coupled with other equations to model processes such as membrane formation \citep{Zhou2006}. The numerical solution is also challenging as the simulations tend to diverge at certain parameter values \citep{Inguva2020, Brun1998}. The problem statement with derivation and assumptions is detailed in the course notes and students are given a brief outline of the problem itself with some examples of polymer blend demixing. While the following derivation is important to develop an understanding of how to simplify complex set of equations, students are not expected to follow this step by step. Emphasis is placed on a conceptual understanding of the Cahn--Hilliard equation and implementation and discussion of solution from FiPy as well as processing using ParaView. 

We consider the case of a binary polymer blend consisting of species A and B. Polymers A and B have a chain length of $N_{A}$ and $N_{B}$, respectively, and the interaction between the two species is characterised by the Flory-Huggins parameter $\chi_{AB}$. As a first set of parameters, students are provided the following values: $N_{A} = N_{B} = 1000$ and $\chi_{AB} = 0.006$. The initial volume fraction of species A, $a_{0}$ was set to $0.77$ with random noise of $\pm 0.03$. The Cahn--Hilliard equation is a fourth order PDE that is typically split and solved as a set of coupled second order PDEs as such \citep{Joki2016}: 
\begin{equation}
\tilde{\mu}_{AB} = \frac{\partial g}{\partial a} - \tilde{\kappa} \tilde{\nabla}^{2}a, 
\end{equation}
\begin{equation}  
\frac{\partial a}{\partial \tilde{t}} = \tilde{\nabla}\cdot (a(1-a)\tilde{\nabla}\tilde{\mu}_{AB}),
\end{equation}
where $\tilde{\mu}_{AB}$ is the non-dimensional difference in chemical potentials between species A and B, $\tilde{t}$ is the non-dimensional time, $a$ is the volume fraction of species A, $g$ is the homogeneous Gibbs free energy of mixing and $\kappa$ is the gradient energy parameter. The equations are scaled with the following time and length scalings: 
\begin{equation}
    t = \frac{R_{G}^{2}}{D_{AB}} \tilde{t}, 
\end{equation}
\begin{equation}
    \mathbf{x} = R_{G} \tilde{\mathbf{x}},
\end{equation}
where $R_{G}$ is the radius of gyration for the polymer and $D_{AB}$ is the diffusion coefficient. 

For polymer blends, $g$ is well represented by the Flory--Huggins equation:  
\begin{equation}
    g = \frac{a}{N_{A}} \ln{a} + \frac{(1-a)}{N_{B}} \ln{(1-a)} + \chi_{AB}a(1-a),
\end{equation}
and $\kappa$ is given as follows \citep{Ariy1990}: 
\begin{equation}
    \tilde{\kappa} = \frac{2}{3} \chi_{AB}.
\end{equation}

\subsubsection{Exercise 4: Results}
\begin{figure*}[htbp]
    \centering
    \begin{subfigure}[t]{0.45\linewidth}
		\centering
		\includegraphics[height=8cm, width=9cm]{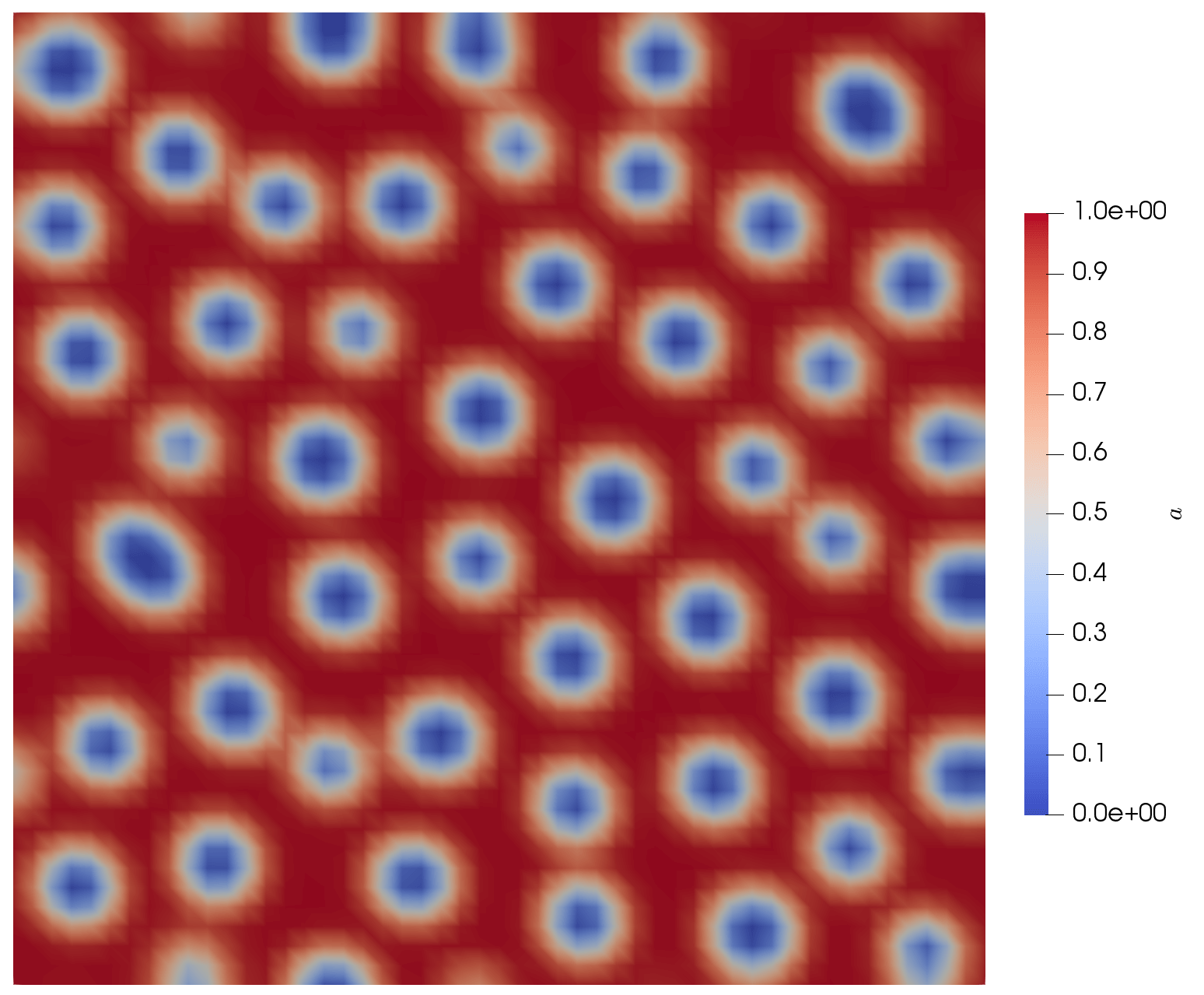}
		\caption{Polymer blend morphology @ $\tilde{t}=20000$. The volume fraction colour bar is included for reference.}
	\end{subfigure}
	\quad
	\begin{subfigure}[t]{0.45\linewidth}
		\centering
		\includegraphics[height=9cm, width=9cm]{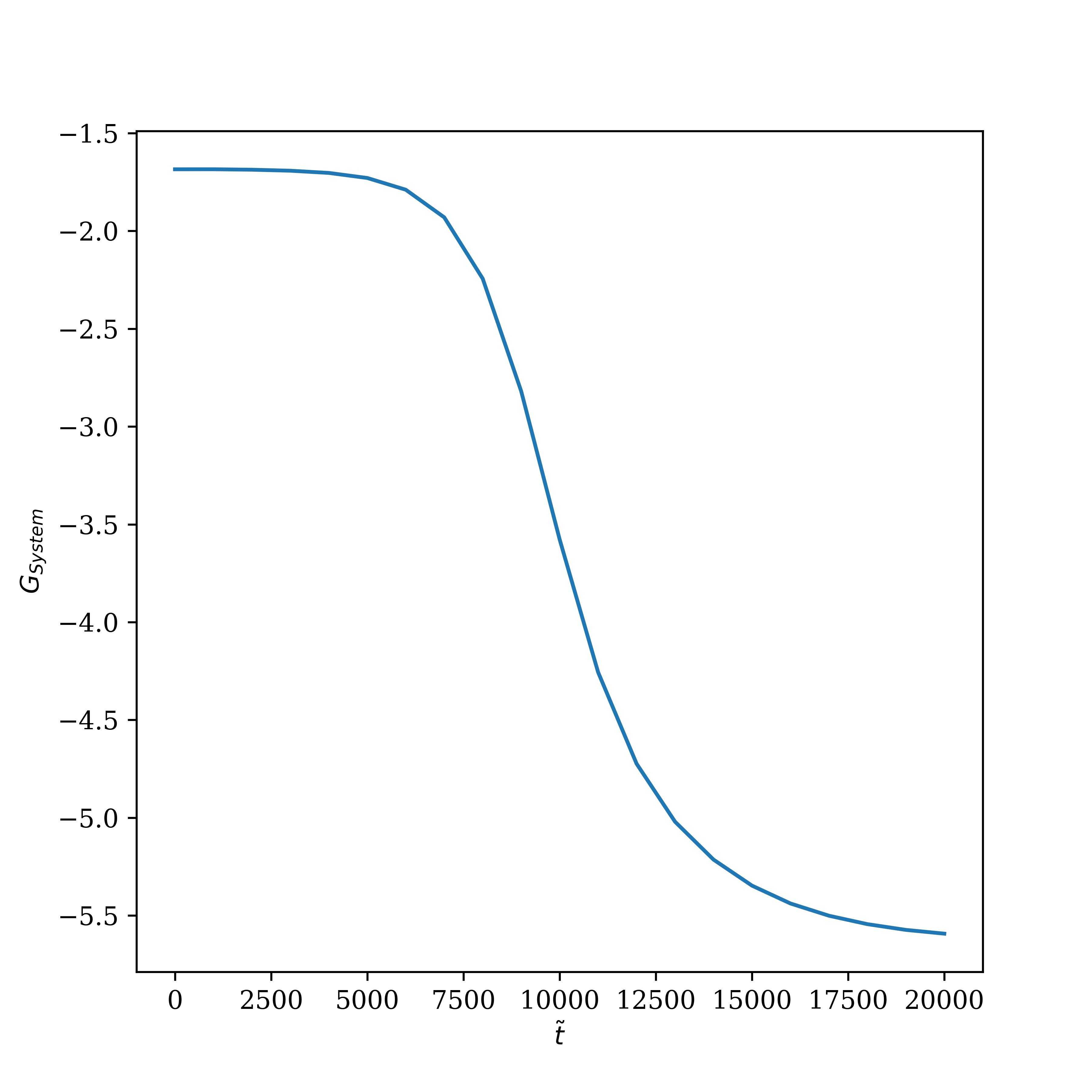}
		\caption{Evolution of $G_{\text{System}}$ over time, demonstrating how the demixing process decreases the total Gibbs energy.}
	\end{subfigure}
	\quad
	\caption{Example figures generated from Exercise 4.}
	\label{fig:ex4_gibbs}
\end{figure*}

While it is possible to use the default FiPy Viewer to view the simulation results, the use of ParaView enables students to explore more advanced post-processing techniques. The system Gibbs free energy $G_{\text{System}}$ as given by \cref{eq:gibbs2component} was evaluated at multiple time-steps in ParaView and is presented in \cref{fig:ex4_gibbs}. It is evident that the Gibbs energy has started to plateau. However, if students continue to let the simulation run, they would be able to observe Ostwald ripening where the smaller blue globules in \cref{fig:ex4_gibbs} "disappear" while the larger globules increase in size while $G_{\text{System}}$ continues to decrease. 

\begin{equation}
    \label{eq:gibbs2component}
    G_{\text{System}}= \int_{V} g(a) + \frac{\kappa}{2}(\tilde{\nabla} a)^2 \ dV. 
\end{equation}

From a numerical perspective, the solution of the modified Cahn--Hilliard equation becomes particularly challenging if students specify excessively large values of $N_{A}$, $N_{A}$ and/or $\chi_{AB}$. For the symmetric system where $N_{A} = N_{B}$, typically the simulation diverges at $N_{A}\chi_{AB} \geq \sim 6$. It must be noted that the development of stable schemes for the numerical solution of the Cahn--Hilliard equations is still an active area of research \citep{Chen2019}. 

\section{Results and discussions}

\subsection{Student feedback and experience}

A total of 9 students signed up for the remote course. We delivered two surveys (referred to as pre-course survey and post-course survey) to assess the change in student self-efficacy and interest in using Python to solve PDEs. In the post-course survey, we also asked about various aspects of design and delivery of the course. In addition to the surveys, we also held focus group at the end of each session of the course to get further details about the content. The analysis of student feedback and survey data is broken down into three categories: Introduction and code familiarisation, content and structure and lastly course delivery and integration. 

\subsubsection{Introduction and code familiarisation}

\begin{figure}[htbp]
    \centering
    \includegraphics[width=1\linewidth]{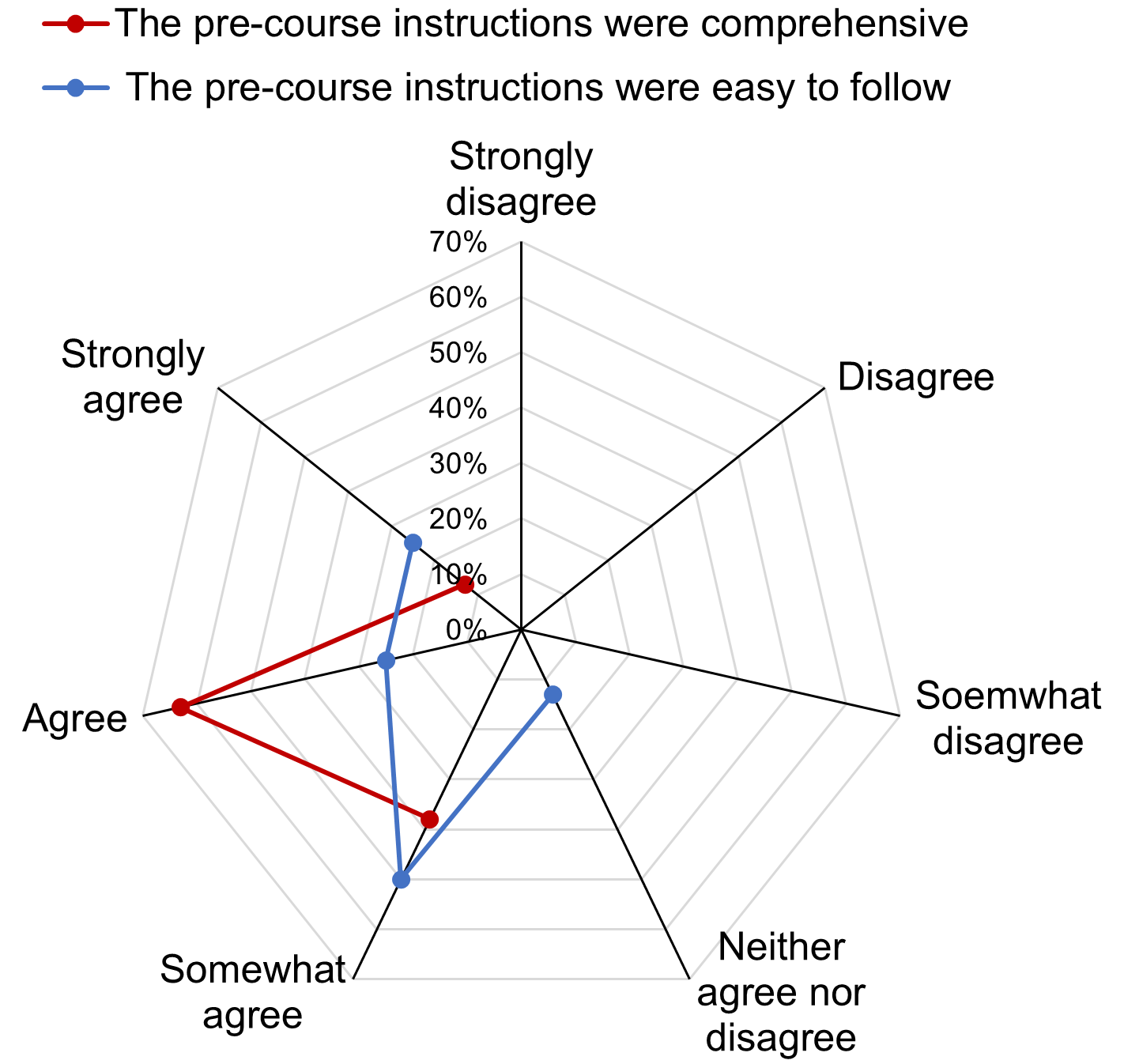}
    \caption{Student response to questions related to pre-course instructions and step-by-step guide for setting up FiPy environment.}
    \label{fig:survey1}
\end{figure}

The development team took significant care to ensure that the set-up process was as smooth as possible. This involved selecting Anaconda and the Spyder IDE as the environment for students to run code, compiling extensive notes with annotated screenshots for set-up. The student response to comprehensiveness and ease to follow these instructional documents is shown in \cref{fig:survey1}. While most of the students agreed that the instructions were comprehensive and easy to follow, there were some students who struggled to set up the environment. This was due to older versions of Anaconda which required uninstalling and installing the latest version to avoid errors. To ensure proper set up, the development team held multiple drop-in sessions for those who were facing issues. 

 Additionally, the first hour of the course was set aside to help student resolve their set-up issues. A common theme in the feedback provided was that more explicit instructions were desired. One student even gave a suggestion of preparing a step-by-step video of the set-up process together with additional information on troubleshooting, indicating it might be more engaging for students compared to written notes.  

\emph{``If you just had that as like a video and then if people get stuck there can be like Oh OK, so the video tells me I just need to do this so...'' }

\begin{figure}[htbp]
    \centering
    \includegraphics[width=1\linewidth]{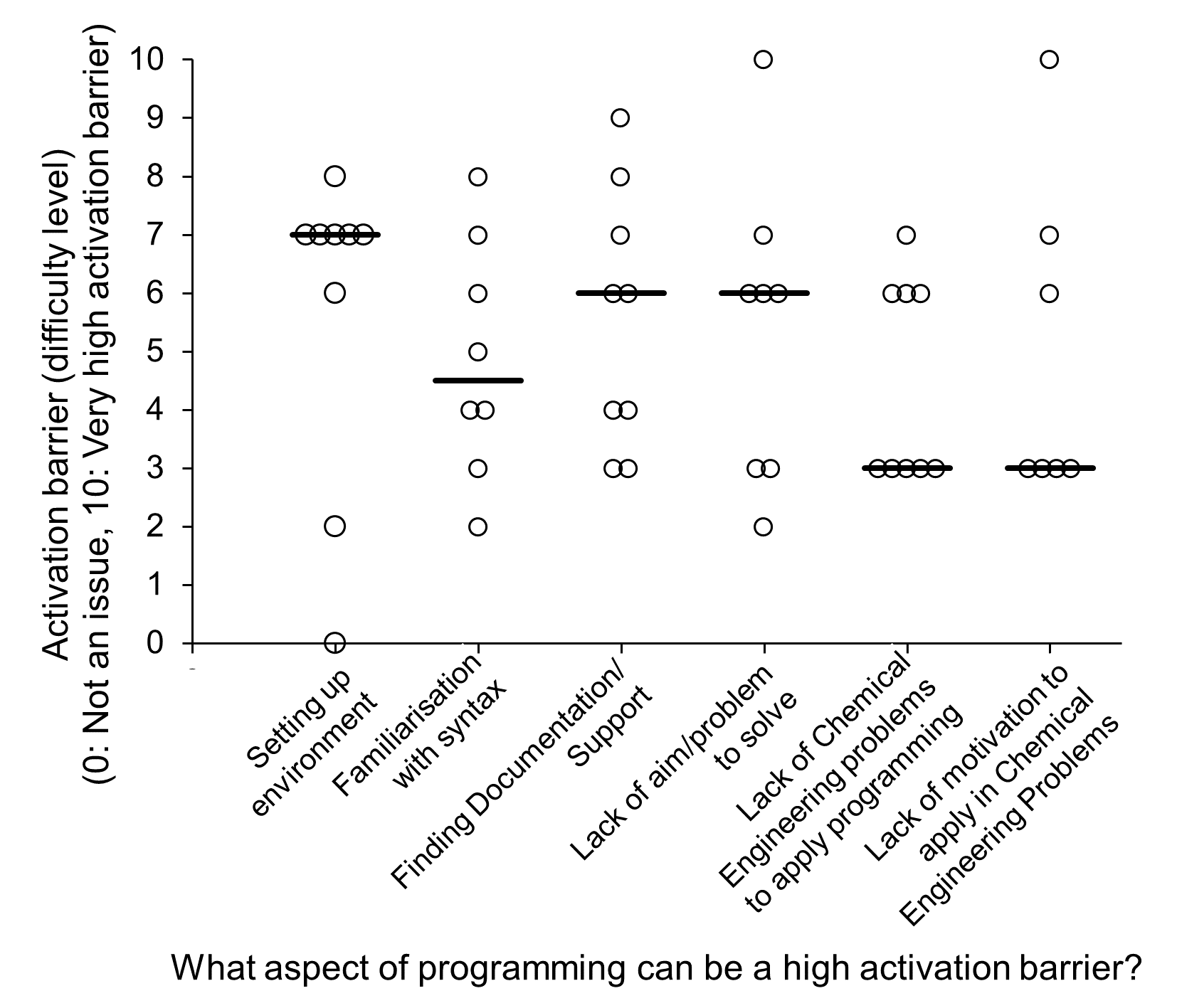}
    \caption{Student rating to aspects of programming with high activation barrier. 0: Not an issue, 10: Very high activation barrier. Solid line represent median.}
    \label{fig:survey2}
\end{figure}

This feedback served as a reminder that even though we had designed the course material to be comprehensive and accessible by our own standards as both educators and users of these research codes, there is always room for improvement. Almost everyone who starts coding will inevitably get frustrated with setting up certain things e.g. an important piece of software or a dependency or even with writing basic code. While such experiences may be instructive, they can be particularly demotivating to beginners. We asked students to rate the activation barriers/difficulties associated with starting and applying Python and the results are shown in \cref{fig:survey2}. Environment set up, familiarization with syntax, and lac of aim/problem to solve posed high activation barriers to programming according to majority of the students. Therefore, due care is needed to reduce this friction as much as possible to create an enabling environment for experiential learning \citep{Inguva2018,Shah2020}.

As most participants did not have any experience in Python, minimal emphasis was placed on syntax with students advised to treat the code as boilerplate, i.e., default codes which they can copy and paste appropriately when needed. We also included the necessary syntax that students would need to be able to adapt the solution as comments that students could comment in or out as desired. This approach worked quite well with most students being able to execute code and make small modifications independently. However, students were generally keen to learn more about Python and how FiPy works and were suggesting that the inclusion of a crash course / refresher in Python might be helpful. 

\emph{``If it's just copying and pasting, I think it'll be right. I was more thinking in the future. For example, if I want to use this myself to solve some problems. Then it would then be nice to also be introduced to how this whole thing is structured.''}

\subsubsection{Content and structure}

Students broadly enjoyed the content of the course, with the visualisation aspect of the numerical solutions being the most well received. This is not a surprising result as providing a visual component can greatly facilitate understanding of concepts, especially more abstract topics such as advanced mathematics \citep{Guzman2002}. An exemplar student comment regarding this point is shown below. 

\emph{``I felt the course explained PDEs better than we got taught it, because you can see it as it was happening and see how one thing affects the other. The graphs gave a more visual approach than when it was taught..."}

We were pleasantly surprised when students demonstrated that they were able to keep up with the pace of the course, especially on the second day when more theoretical and complex aspects of numerical methods, PDE theory and thermodynamics related to exercises three and four were introduced. Students broadly acknowledged that the course structure was logical and facilitated their progression through the course material. The further division of exercises two and three into distinct parts with clear goals also helped students by making clear the learning point for that particular part of the exercise. 

\emph{``I like the fact that you separated into like parts ABC so people could see which code coincided with which sort of change in like the equation that we're making...''}

\begin{figure*}[htbp]
    \centering
    \begin{subfigure}[t]{0.45\linewidth}
		\centering
		\includegraphics[height=7cm, width=8cm]{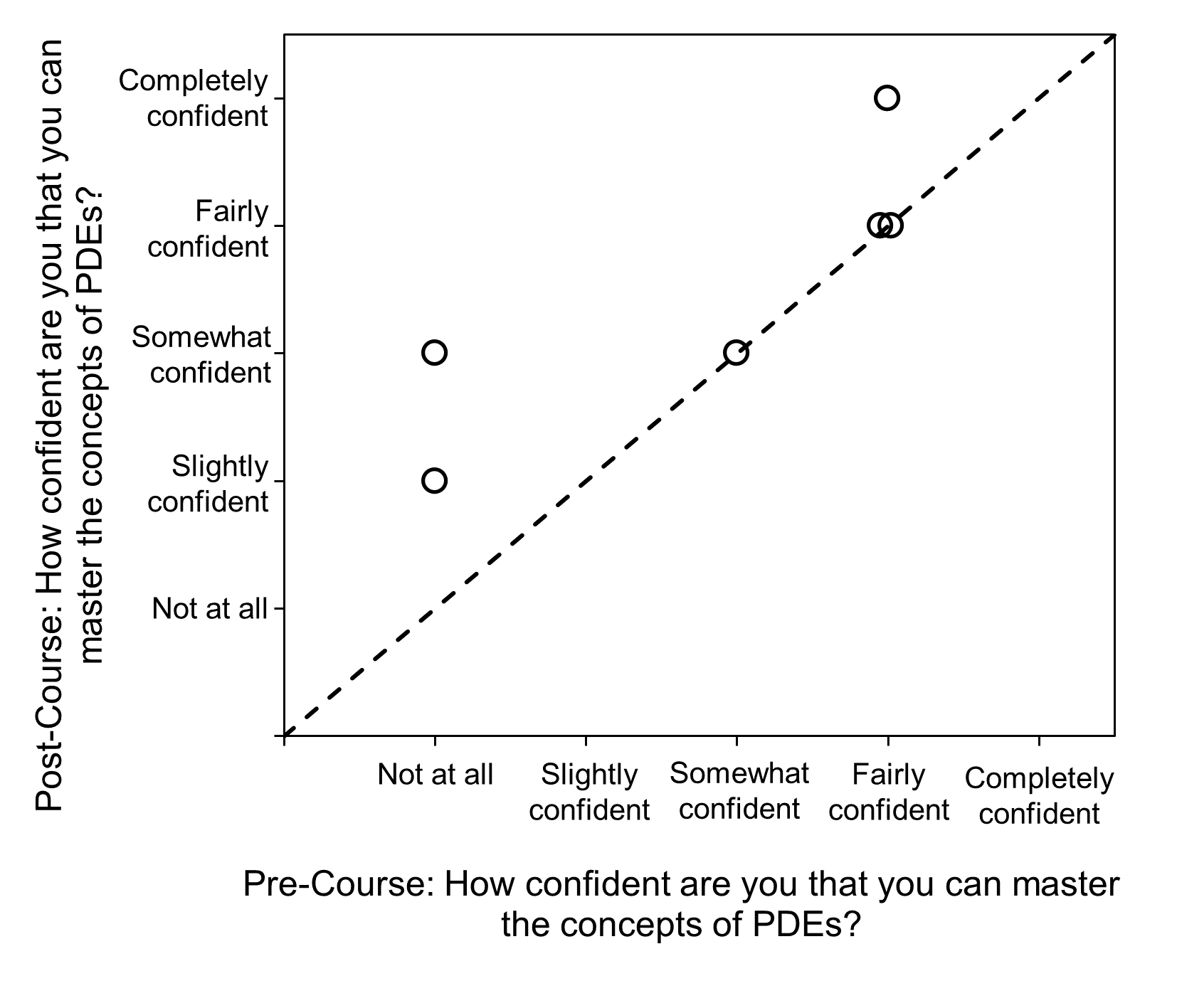}
		\caption{Student response to their self-perceived confidence to master the concepts of PDE.}
		\label{fig:survey3_MPVP}
	\end{subfigure}
	\quad
	\begin{subfigure}[t]{0.45\linewidth}
		\centering
		\includegraphics[height=7cm, width=8cm]{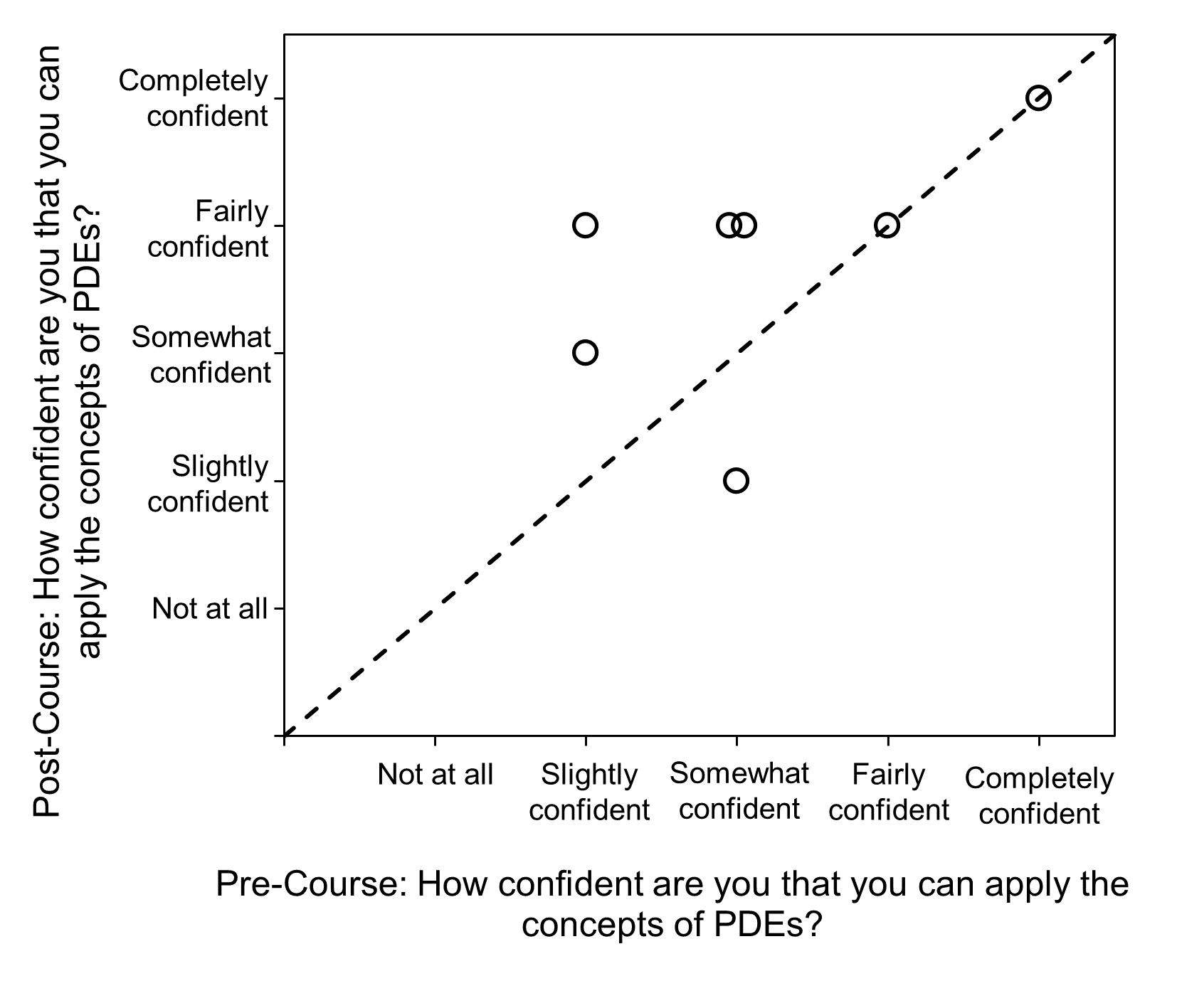}
		\caption{Student response to their self-perceived confidence to apply the concepts of PDE.}
		\label{fig:survey3_APVP}
	\end{subfigure}
	\quad
	\begin{subfigure}[t]{0.45\linewidth}
		\centering
		\includegraphics[height=7cm, width=8cm]{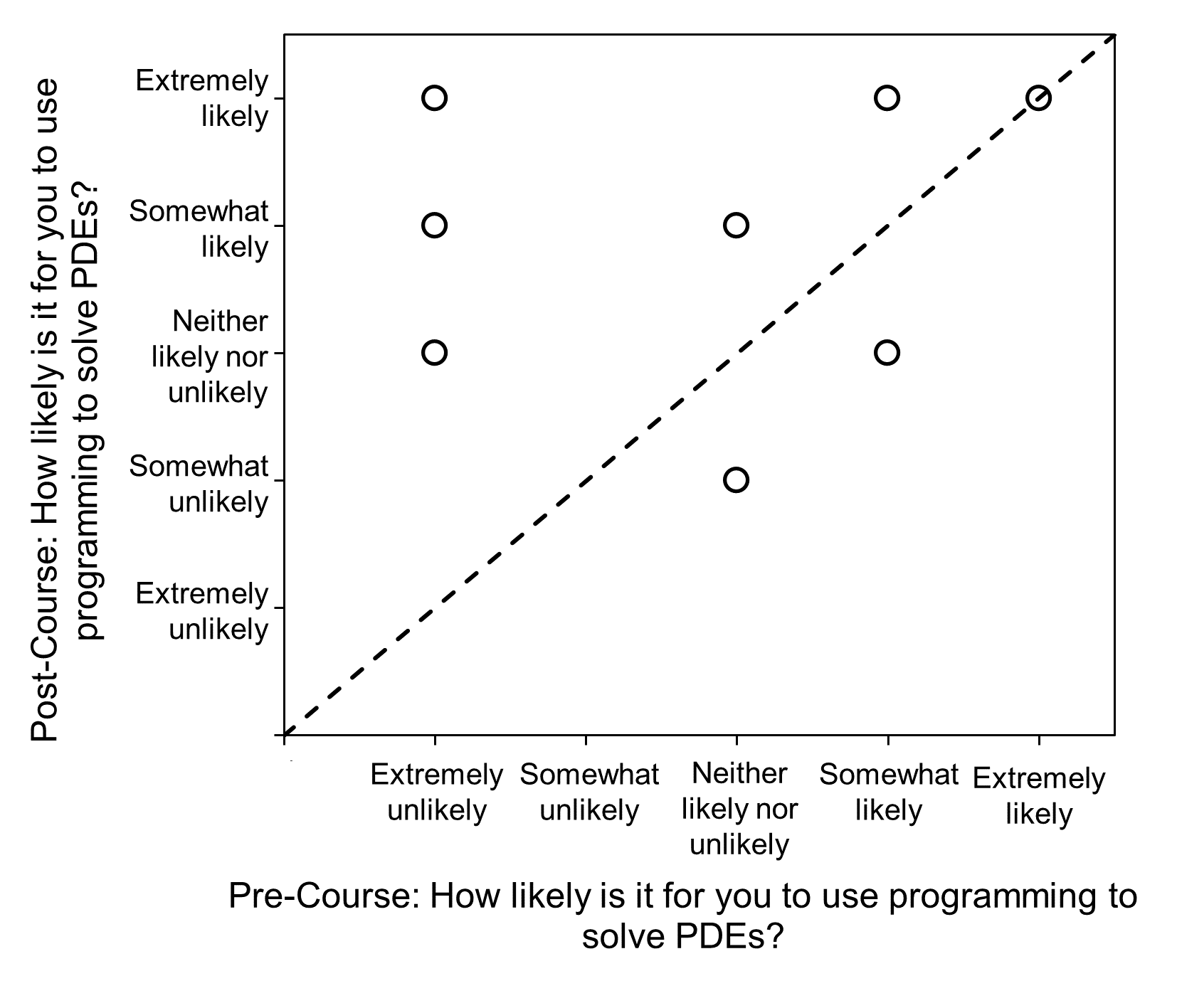}
		\caption{Student response to their likelihood of using programming to solve PDEs (pre course vs. post course).}
		\label{fig:survey3_LP}
	\end{subfigure}
	\quad
	\begin{subfigure}[t]{0.45\linewidth}
		\centering
		\includegraphics[height=7cm, width=8cm]{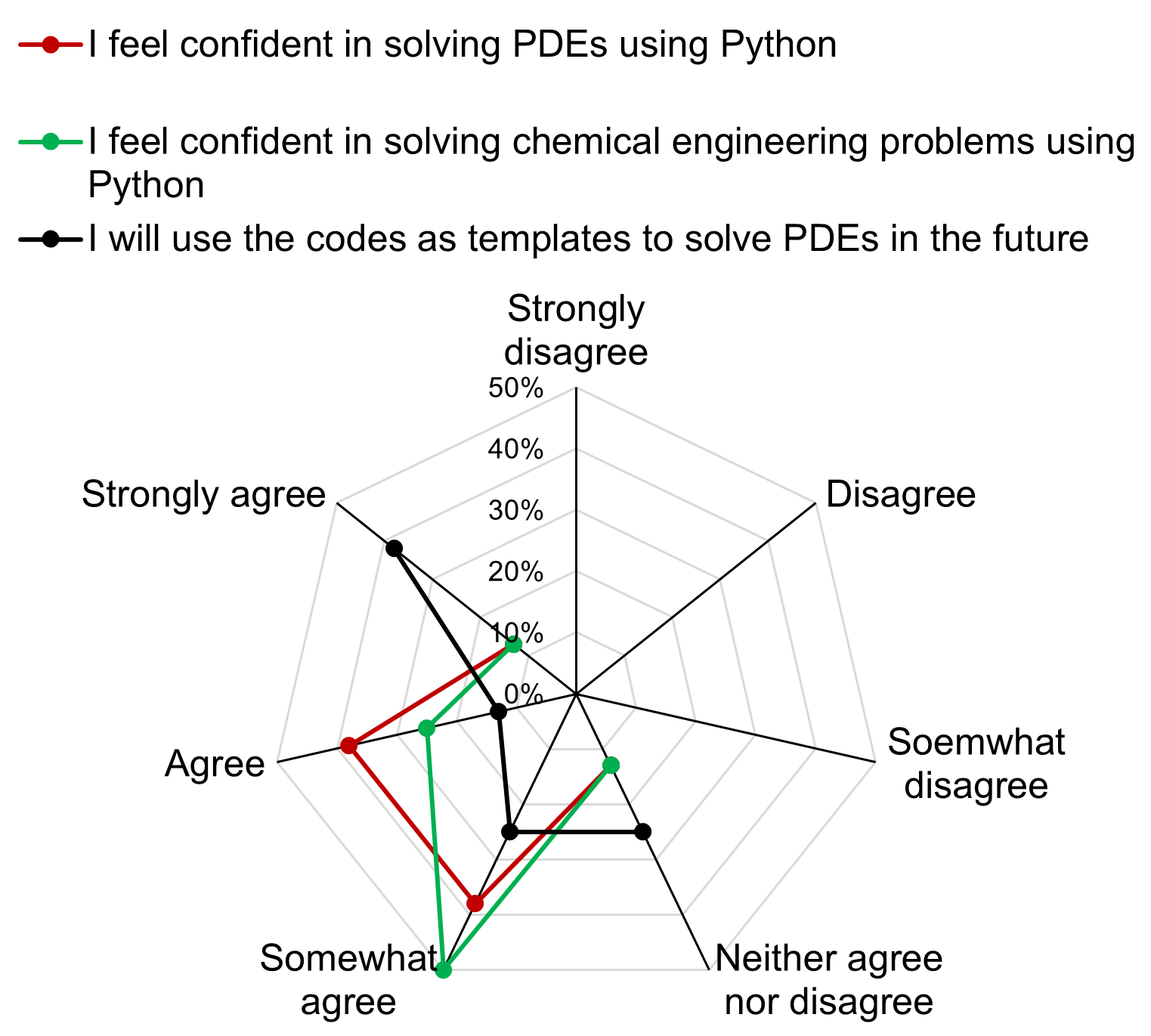}
		\caption{Student responses to their confidence in solving problems and using Python in the future after the course.}
		\label{fig:survey3_FU}
	\end{subfigure}
	\quad
	\caption{Student responses to questions related to their self-efficacy and effectiveness of the course.}
	\label{fig:survey3}
\end{figure*}

We asked student questions related to their self-efficacy including their self-perceived confidence to a) master and b) apply concepts of PDEs in the pre-course and post-course surveys. The results are shown in \cref{fig:survey3}. Students who expressed no confidence or low confidence in the pre-course survey showed higher confidence in the post-course survey suggesting the course improved their self-efficacy related to PDEs. The course had most benefit in perception of students who were extremely unlikely to use programming to solve PDEs. One of the key objectives of the course was to provide a starting point for students to use Python for solving PDEs and several students responded favorably to using the codes as templates in the future (\cref{fig:survey3}). 

\subsubsection{Course delivery and integration}

Due to the current COVID-19 pandemic, the course had to be run remotely. Fortunately, due to the various features made available in online collaboration platforms such as Microsoft Teams, the course could be delivered smoothly with no students indicating that remote delivery adversely impacted their experience. One way Microsoft Teams facilitated this process was through the feature of being able to create polls in the chat function. This enabled teaching staff to create polls at specific intervals to gauge students` understanding of a specific point or to gather information on what aspects of the code did students independently explore during breaks. 

Most students indicated a desire for more active and structured learning components related to the actual code and the running of simulations during the course. The original design of the course centered around providing students the flexibility to independently explore various aspects of each exercise such as different boundary and initial conditions or different numerical schemes with the exercise script serving as a basis for students to then modify. This approach was not favoured by students. Many of the suggestions regarding this point indicated that students desired a couple of formal and structured questions or tasks which they could execute at the end of each exercise and compare to a correct solution to consolidate their learning. We intend to implement this as part of the second iteration of the course. 

\emph{``If you had a couple of exercise questions or stretch questions for your knowledge at the end and you were like we would like you to obtain this graph or we would like you to change this parameter and think about the understanding that it has or model this question(from the problem sheets or exams) ... Something like that, and then you can have a predictable outcome that people can reproduce with just a few exercise questions at the bottom I think would be helpful.''}

Regarding the integration of the activities from the course into lecture based courses such as the Math II module, the first and most readily adaptable aspect is the generation of high quality plots and animations for visualisation purposes. Beyond that, the inherently open-ended nature of such a tool lends itself easily to being used as the basis of a coursework which could be based on a specific example from the range of equation systems covered in the present work or even a new system that could be of interest to students.

\section{Conclusions}

We have demonstrated how the PDE solver code FiPy can be effectively introduced to students, with very little coding experience, in a short period of time. The key to this is designing the course to be as accessible as possible and designing each exercise to build on previously taught material, enabling faster progression. Referencing the four factors outlined previously for evaluating codes for use in the classroom, FiPy and other PDE solver codes in Python do well in all four aspects. The open-source nature of  all the tools used in the present work: Anaconda, FiPy and ParaView, mean that availability is not an issue. These tools are also frequently used in industry and academia, demonstrating their scalability. Lastly, we have demonstrated how a variety of problems of different complexities can be easily implemented and rolled-out to students, highlighting the ease-of-use of these tools and the ability for them to be tailored to meet a variety of learning  objectives. This course improved student perception to Python and their self-efficacy in concepts of PDEs with students with less confidence showing the greatest boost. 

The present work also reveals various multiple challenges and issues educators should take note of when rolling out such an introductory course. Despite our best efforts to make the set-up process as seamless as possible, students indicated this still remained a significant pain point. Fortunately, receiving such student input at a small scale course means that this can be addressed in future iterations. Regarding the structure of the course, educators should provide more structured tasks during the content delivery process so that students can consolidate their learning and remain engaged. These tasks could take the form of simple tasks such as modifying the initial/ boundary conditions of a problem, changing values of certain dimensionless parameters, and comparing it to a model solution. 

The highly flexible nature of these codes enables them to be used in a variety of settings beyond the introductory course we have presented. A review of the various user discussion forums or code documentation reveals several more capabilities and use cases that can be integrated into the classroom in a variety of topics at different academic levels. We hope to explore the nuances between the different codes and discretisation schemes and its integration into a more advanced course of numerical methods in future work. 
\section{Acknowledgements}

Authors would like to thank students who participated in this course and provided their valuable feedback.

\section{Availability of code}
All the course notes and code can be found at the following repository: https://github.com/pavaninguva/pde-Solver-Course

\bibliography{references}

\end{document}